\documentclass[draftcls, onecolumn,12pt]{IEEEtran}

\usepackage{amsmath}
\usepackage{cases}
\usepackage{amsfonts}
\usepackage{amssymb}
\usepackage{boxedminipage}
\usepackage{graphicx}
\usepackage[usenames]{color}
\usepackage{array,color}
\usepackage{colortbl}
\usepackage{lineno}
\usepackage{slashbox,pict2e}

\pagewiselinenumbers

\begin{document}

 \baselineskip=22pt
\title{Service Based High-Speed Railway Base Station Arrangement}
\author{Chuang~Zhang$^\ast$,~Pingyi~Fan$^\ast$,~Yunquan~Dong$^\ast$,  Ke~Xiong$^\ast$$^\dag$\\  

$^\ast$Department of Electronic Engineering, Tsinghua University, Beijing, P.R. China\\
$^\dag$School of Computer and Information Technology, Beijing Jiaotong University, Beijing, P.R. China\\
E-mail:~\{zhangchuang11,~dongyq08\}@mails.tsinghua.edu.cn,\{fpy,~kxiong\}@tsinghua.edu.cn}
\maketitle

\begin{abstract}
To provide stable and high data rate wireless access for passengers in the train, it is necessary to properly deploy base stations along the railway. We consider this issue from the perspective of service, which is defined as the integral of the time-varying instantaneous channel capacity. With large-scale fading assumption, it will be shown that the total service of each base station is inversely proportional to the velocity of the train. Besides, we find that if the ratio of the service provided by a base station in its service region to its total service is given, the base station interval (i.e. the distance between two adjacent base stations) is a constant regardless of the velocity of the train. On the other hand, if a certain amount of service is required, the interval will increase with the velocity of the train. The above results apply not only to simple curve rails, like line rail and arc rail, but also to any irregular curve rail, provided that the train is travelling at a constant velocity. Furthermore, the new developed results are applied to analyze the on-off transmission strategy of base stations.
\end{abstract}


\section{Introduction}

The rapid growth of high-speed railways around the world brings huge demands for broadband wireless communications on high-speed trains, however, the current dominant system GSM for railway (GSM-R) can only support data rate less than 200kbps, and is mainly used for signalling rather than data transmission \cite{wang2012dasmchst}. Recently, some broadband wireless communication systems for high speed trains have been developed, for instance, combining satellite and terrestrial methods, Thalys in Europe can provide passengers broadband Internet access service. The trains Shinkansen N700 in Japan can guarantee users 2Mbps bandwidth via leakage cable. However, widespread application of these systems are quite limited since they are costly and much dedicated \cite{barbu2010}. Therefore, it is necessary to develop new technologies and systems to meet the needs of high-speed railway wireless communication. Either simply making amendments to GSM-R or evolving the current system to a more advanced one requires the solution of several fundamental issues.

Firstly, radio signals experience dramatic loss when penetrating into or from the alloy carriage, which imposes a high energy burden on the transmitter. Secondly, severe Doppler shift exists in the communication process, for instance, the Doppler shift could be 833Hz if the train travels at 300km/h and the carrier frequency is 3GHz, this would result in the difficulties of synchronization \cite{yang2012dfoe}. Thirdly, frequent handovers due to high mobility could lead to the increase of drop-offs, which significantly degrades user experiences. Fourthly, channel measurement is difficult as a result of diverse transmission scenarios like viaducts, tunnels, and open fields and severe Doppler frequency shift. Finally, in order to guarantee a stable service, it is important to properly deploy base stations along the railway.

Many works have aimed at solving these problems. For instance, to deal with the first problem, a two-hop architecture was proposed in \cite{barbu2010} \cite{lin2002cehspts}. In the system, user information is firstly transmitted to the access point (AP) in the train, then the AP sends the aggregated information to the base station via an antenna on the top of the train. In this way, direct transmission from user terminals to the base station is avoided. The performance difference of this structure versus direct transmission was analyzed in \cite{you2012tshsrdr}. As for the second problem, several Doppler spread compensation methods such as the REM-based algorithm in \cite{li2012remcd} and joint Doppler frequency shift compensation and data detection method using 2-D unitary ESPRIT algorithm in \cite{yi2012jdfscddm} were developed. Concerning the third problem, a Radio-over-Fiber distributed antenna system was proposed in \cite{bart2007rofbs}. In this system, antennas are allocated along the railway and connected to a control center through fiber to increase the coverage region of a base station, thereby leading to the reduction of handover frequency. The coverage efficiency of this system was analyzed in \cite{zhang2010cerofn}.  Besides, several handover schemes were proposed for the high-speed railway scenario in \cite{yamada2010csfho} \cite{lin2002sdlhos} \cite{li2012ahots} \cite{cheng2012bphos}. For the fourth problem, channel fading statistics in high-speed mobile environment were studied in \cite{wen2012cfshsm}, and some channel measurements methods were reviewed in \cite{zhou2012hsrcmc}.  Finally, regarding the fifth issue, signal strength is usually the key factor to the base station arrangement, i.e. to guarantee the received signal to noise ratio higher than a given threshold in the boundary of two base stations. From an information-theoretic point of view, this guarantees the instantaneous channel capacity to be larger than a threshold. However, if the train moves very swiftly, like the high-speed railway case, the instantaneous channel capacity changes dramatically with the movement of the train, and each base station could only maintain its service to the train for a short period. For instance, if a base station covers  3km along the track, it can provide service for a train with velocity 300km/h for  36 seconds. After these 36 seconds, the train has to connect to a new base station for Internet access. Therefore, it would be valuable to consider the transmission of the base station in that period as a whole (i.e. what the base station can provide in that period rather than at a time point). Using the service requirement for a base station, i.e. each base station is required to provide a certain ratio or amount of service for the moving train in its service region, we can decide the service region of each base station, and if we assume service regions of all base stations are disjoint, the base station interval can be determined correspondingly.

In this work, we discuss base station arrangement on three different scenarios: the line rail, the arc rail and the irregular curve rail. The first two kinds of rails are typical ones since they are most common in the practical environment. For the line rail case, we develop an explicit form of the relationship between the base station interval and the velocity of the train given service requirement. In particular, we find that when using the ratio of the total service to decide the interval, the interval is a constant regardless of the velocity of the train. On the other hand, if a base station is required to provide a given amount of service (less than the total service), the interval will increase with the velocity of the train. For the arc rail, we use the central angle of the arc between two base stations rather than the base station interval to decide the position of each base station and find that the relationship between the angle and the velocity of the train is similar to the relationship between the interval and the velocity. For the irregular curve rail, we will prove that the service of each base station between any two points is also inversely proportional to the velocity of the train. In addition, for the irregular curve, we provide two base station arrangement algorithms when the variation of the curve is not large. Finally, we apply the main results to analyze the on-off transmission strategy of base stations.

The rest of this paper is organized as follows. Definitions and system model are introduced in Section \ref{sec_systmod}. Detailed discussions about base station arrangement for the line rail are given in Section \ref{sec_disobsi}. In Section \ref{sec_arcrail} and Section \ref{sec_anycurverail}, we discuss the deployment of base stations along an arc rail and an irregular curve rail,  respectively. In Section \ref{sec_scrttr}, the main results are applied in the analysis of on-off transmission strategy of base stations. Finally, conclusions are given in Section \ref{sec_conclusion}.

\section{System Model}\label{sec_systmod}

\subsection{System Architecture}

\begin{figure}
  \centering
  \includegraphics[width=0.48\textwidth]{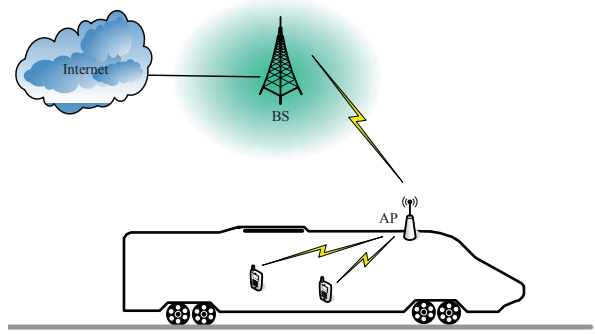}\\
  \caption{Two-hop architecture.}\label{fig_twohop}
\end{figure}
We consider a two-hop transmission model as shown in Fig. \ref{fig_twohop}. To access the network, users in the train firstly connect to the access point (AP) in the carriage, the AP then forwards the user data to base stations via the antenna on the top of the train. The downlink from base stations to user terminals reverse this process. Such a structure has several advantages. Firstly, the drop-off rate could be reduced significantly since the base station only needs to deal with the handoff of one AP rather than dozens of user terminals. Secondly, using the antenna on the top of the train, signals don't penetrate into or from the carriage, thus avoiding heavy energy loss. Finally, some well developed technologies like Wifi can be used in the carriage to provide a stable and high speed wireless link.

In the downlink of this two-hop architecture, the first hop (i.e. the base station to AP) plays a key role in the service of base stations, and may become the bottleneck in this architecture. Therefore, we mainly discuss the base-station-to-AP transmission and employ the point to point communication model for this part.

Since the effect of large-scale fading is more obvious than that of small-scale fading (especially in open fields where most of China's railways are laid), we ignore the channel variation due to small-scale fading and assume that the change of received signal strength results only from the position shifts of trains. So we can express the path loss as a function of the distance between the base station and the train: $L_{p}(d(t))=A+10\alpha\log(d(t))$, where $A$ is a factor affected by the carrier frequency, the heights of the transmitter and receiver antennas, different climate or geology conditions etc, $\alpha$ is the path loss exponent and $\alpha\geq 2$, $d(t)$ is the distance between the transmitter and the receiver at time $t$. In our later discussions, we assume that $A$ is a constant during the transmission, in this case, we can normalize it without affecting the analysis, and the path loss can be expressed as  $L_{p}(d(t))=d^{\alpha}(t)$.

Assuming that the base station and the AP communicate over an additive white Gaussian noise(AWGN) channel. At the moment $t$, the channel output is
\begin{equation}\label{equ_awgnout}
  Y_t=\frac{1}{\sqrt{L_p(d(t))}}X_t+Z_t,
\end{equation}
where $Z_t\sim N(0,N_0/2)$ and independent of channel input $X_t$.
Given a power constraint $P_s$, the capacity of the discrete-time AWGN channel \cite{abbas2011nit} is:
\begin{equation}\label{equ_awgncap}
  C(P_s)=\log(1+\frac{P_s}{L_p(d(t))N_0/2}).
\end{equation}
$\frac{P_s}{L_p(d(t))N_0/2}$ is the received signal to noise ratio (SNR).

\begin{figure}
  \centering
  \includegraphics[width=0.48\textwidth]{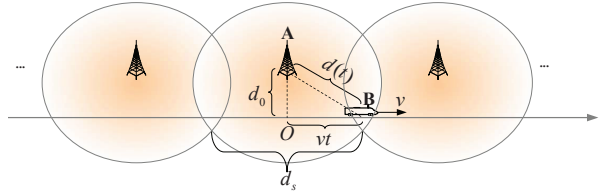}\\
  \caption{Base station coverage model.}\label{fig_bscover}
\end{figure}

The base station coverage model is given in Fig. \ref{fig_bscover}. In this figure, the base station is represented as A, the train is denoted as B. Assuming that the train is moving at a constant velocity $v$ along the railway, the base station with $d_0$ meters from the railway transmits signals to the train using constant power $P_s$. Besides, for convenience, we establish an axis along the railway: Let the cross point between the railway and its vertical line through the base station be the original point, the moving direction of the train be the positive direction, the instant when the train passes the origin be time $0$. Then at a given time $t$ ($t\in[-\infty,\infty]$), the distance $d(t)$ is $\sqrt{d_0^2+(vt)^2}$. The received SNR at time $t$ is
\begin{eqnarray}\label{equ_snr}
  SNR &=&\frac{\frac{P_s}{(d_0^2+(vt)^2)^{\frac{\alpha}{2}}}}{N_0/2}\nonumber \\
  &=&\frac{2P_s}{(d_0^2+(vt)^2)^{\frac{\alpha}{2}}\,N_0}.
\end{eqnarray}

Substituting Eqn. (\ref{equ_snr}) into Eqn. (\ref{equ_awgncap}), we get the instantaneous channel capacity at time $t$,
\begin{equation}\label{equ_capaomt}
  C(t)=\log(1+\frac{2P_s}{(d_0^2+(vt)^2)^{\frac{\alpha}{2}}\,N_0}).
\end{equation}

For convenience, let $\gamma$ represent $\frac{2P_s}{N_0}$, which is the transmitter SNR, then Eqn. (\ref{equ_capaomt}) can be rewritten as
\begin{equation}\label{equ_capaomts}
  C(t)=\log(1+\frac{\gamma}{(d_0^2+(vt)^2)^{\frac{\alpha}{2}}})
\end{equation}

\begin{figure}
  \centering
  \includegraphics[width=0.48\textwidth]{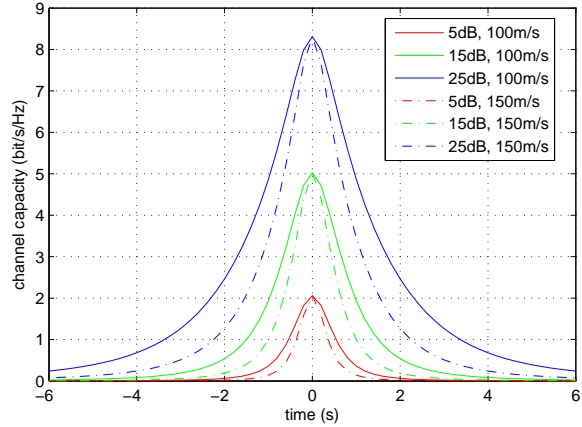}\\
  \caption{Instantaneous channel capacity versus time for $SNR_0=5,15,25 dB$, $v=100m/s,150m/s$, $d_0=50m$, $\alpha=3$.}\label{fig_capatim}
\end{figure}

The relationship between instantaneous channel capacity and time is plotted in Fig. \ref{fig_capatim}. The $SNR_0$ denotes the received SNR at point $O$, i.e. $SNR_0=\frac{\gamma}{d_{0}^{\alpha}}$. It is observed that $i)$ the channel capacity tends to zero in both directions, $ii)$ when the $SNR_0$ are the same, the higher the velocity of the train, the smaller the channel capacity at the same time point.

\subsection{Definition of Service}

We define \emph{service} as the integral of the instantaneous channel capacity over a given time period, similar as in \cite{dong2012spfmc}\cite{dong2012dtlspfc}. It upper bounds the transmission of a base station and represents the maximum amount of data that the physical layer can provide for the network layer. Specifically, it is
\begin{equation}\label{equ_servcap}
  S(t)=\int_{-\infty}^{t}C(\tau)\,\mathrm{d}\tau.
\end{equation}

It is easy to see that $S(-\infty)=0$, and $S(\infty)=\int_{-\infty}^{\infty}C(\tau)\,\mathrm{d}\tau$ which represents the total service of the base station.

Substituting Eqn. (\ref{equ_capaomts}) into Eqn. (\ref{equ_servcap}), we get
\begin{equation}\label{equ_servofmt}
  S(t)=\int_{-\infty}^{t}\log(1+\frac{\gamma}{(d_0^2+(v\tau)^2)^{\frac{\alpha}{2}}})\,\mathrm{d}\tau.
\end{equation}

\begin{figure}
  \centering
  \includegraphics[width=0.48\textwidth]{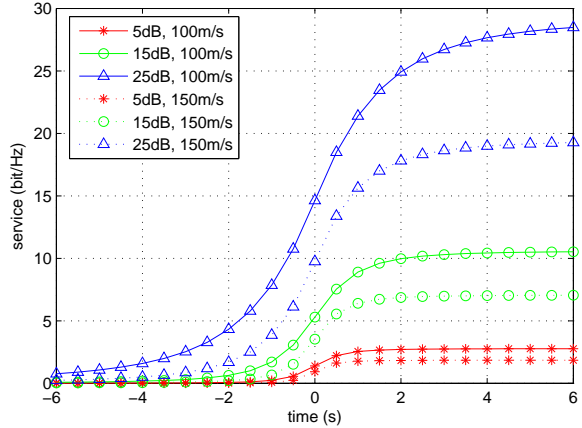}\\
  \caption{Service versus time for $SNR_0=5,15,25 dB$, $v=100m/s,150m/s$, $d_0=50m$, $\alpha=3$.}\label{fig_servtim}
\end{figure}

Curves of service versus time with different $SNR_0$ and velocities are plotted in Fig. \ref{fig_servtim}. Since the service at time $t$ is equal to the integration of the instantaneous channel capacity before $t$, with $t$ sufficiently large, the channel capacity tends to $0$, and the service tends to a constant. Besides, the results also show that the higher the velocity, the smaller the total service.

\subsection{Definition of Service Distance}
Assuming that each base station has a disjoint \emph{service distance} $d_s$, as shown in Fig. \ref{fig_bscover}, and to avoid interference, the station transmits only when the train is within its service distance. Besides, to make the instantaneous channel capacity large, we choose the service distance of each base station symmetrically around it. Thus if the train is in the service region of a base station from time $-t_s$ to $t_s$, we have
\begin{equation}\label{equ_servdis}
  d_s=2vt_s.
\end{equation}

Since trains move along a linear rail, we can lay out the base stations along the track in one-dimension as shown in Fig. \ref{fig_bscover}. If the service distances of two adjacent base stations are $d_{s_1}$ and $d_{s_2}$, respectively, the interval $d_{int}$ of these two base stations is
\begin{equation}\label{equ_interval}
  d_{int}=\frac{d_{s_1}+d_{s_2}}{2}.
\end{equation}

In next section, we consider the case where the service distances of all base stations are the same, then the base station interval equals the service distance, so we will not distinguish these two quantities, and use $d_s$ to denote both of them.

\section{Discussion on the Line Rail}\label{sec_disobsi}

\subsection{ Total Service of a Base Station}
The total service of a base station is the integration of instantaneous channel capacity from $-\infty$ to $\infty$, given by
\begin{equation}\label{equ_toserq}
  S_{tot}=\int_{-\infty}^{\infty}\log(1+\frac{\gamma}{(d_0^2+(v\tau)^2)^{\frac{\alpha}{2}}})\,\mathrm{d}\tau.
\end{equation}
When using the substitution $x=v\tau$ in the above equation, we have
\begin{equation}\label{equ_toseqx}
  S_{tot}=\frac{1}{v}\int_{-\infty}^{\infty}\log(1+\frac{\gamma}{(d_0^2+x^2)^{\frac{\alpha}{2}}})\,\mathrm{d}x.
\end{equation}
As can be seen from the above equation, the total service of a base station is decided by the transmitter SNR $\gamma$, the distance of the base station from the rail $d_0$, the path loss exponent $\alpha$ and the velocity of the train $v$. In particular, when the transmitter SNR, the path loss exponent and the distance are given, the total service is inversely proportional to the velocity of the train.

\subsection{ Choice of Base Station Interval given a Service Ratio Requirement}\label{sec_bsisr}

It is unreasonable to require the base station to provide its total service to a moving train, since the channel capacity would be too small to be effective when the train is far from the base station. Instead, we require that the base station provide a certain ratio ($60\%$ for instance) of its total service to the train using its best capability. Then we can use this ratio to decide the service distance of the base station (correspondingly the base station interval). In fact, it will be shown later that this method is equivalent to deciding the base station interval guaranteeing a minimum channel capacity.

If from time $-t_s$ to $t_s$, the train is in the service distance of a base station, then the ratio of the service the base station can provide to its total service is
\begin{equation}\label{equ_domirat}
  \eta=\frac{\int_{-t_s}^{t_s}C(\tau)\,\mathrm{d}\tau}{\int_{-\infty}^{\infty}C(\tau)\,\mathrm{d}\tau},
\end{equation}
Obviously, $\eta\sim[0,1]$.

Regarding the base station interval decided by the service ratio, we have the following result:

\emph{\textbf{Proposition 1:}} \emph{Given the transmitter SNR $\gamma$, the path loss exponent $\alpha$ and the distance from the base station to the rail $d_0$, the base station interval $d_s$ is uniquely decided by a certain service ratio $\eta$ and irrelevant to the velocity of the train}.

\begin{proof}
Assuming that the train is travelling with a constant velocity $v$, from time $-t_s$ to $t_s$, it is in the service region of the base station $A$ as in Fig. \ref{fig_bscover}, then the service ratio is

\begin{eqnarray}\label{equ_etavsd}
  \eta &=& \frac{\int_{-t_s}^{t_s}\log(1+\frac{\gamma}{(d_0^2+(v\tau)^2)^{\frac{\alpha}{2}}})\,\mathrm{d}\tau}
   {\int_{-\infty}^{\infty}\log(1+\frac{\gamma}{(d_0^2+(v\tau)^2)^{\frac{\alpha}{2}}})\,\mathrm{d}\tau} \nonumber \\
   &=& \frac{\int_{0}^{t_s}\log(1+\frac{\gamma}{(d_0^2+(v\tau)^2)^{\frac{\alpha}{2}}})\,\mathrm{d}\tau}{\int_{0}^{\infty}\log(1+\frac{\gamma}{(d_0^2+(v\tau)^2)^{\frac{\alpha}{2}}})\,\mathrm{d}\tau}.
\end{eqnarray}
This can be rewritten as
\begin{equation} \label{equ_chform}
\textstyle
\int_{0}^{t_s}\log(1+\frac{\gamma}{(d_0^2+(v\tau)^2)^{\frac{\alpha}{2}}})\,\mathrm{d}\tau=\eta\int_{0}^{\infty}\log(1+\frac{\gamma}{(d_0^2+(v\tau)^2)^{\frac{\alpha}{2}}})\,\mathrm{d}\tau.
\end{equation}
Using the substitution $x=v\tau$ for both sides in the above equation, and note that $d_s=2vt_s$, we have
\begin{equation}\label{equ_Cetasim}
  \int_{0}^{\frac{d_s}{2}}\log(1+\frac{\gamma}{(d_0^2+x^2)^{\frac{\alpha}{2}}})\,\mathrm{d}x= \eta\int_{0}^{\infty}\log(1+\frac{\gamma}{(d_0^2+x^2)^{\frac{\alpha}{2}}})\,\mathrm{d}x.
\end{equation}
Given the transmitter SNR $\gamma$, the path loss exponent $\alpha$ and the distance from the base station to the rail $d_0$, the left side of the above equation is a monotonically increasing function of $d_s$, which we denote as $f_1(d_s)$, the integral on the right side is a constant which we denote as $C_1$. So the above equation can be written in a simple form
\begin{equation}
f_1(d_s)=C_1\eta.
\end{equation}
Since $f_1(d_s)$ is invertible, we can get
\begin{equation}\label{equ_dssimp}
  d_s=f_1^{-1}(C_1\eta).
\end{equation}
As can be seen from this equation, when the ratio $\eta$ is given, the base station interval $d_s$ is uniquely determined, which is irrelevant to the velocity of the train.
\end{proof}

According to this proposition, if we use the service ratio to determine the base station interval, the interval $d_s$ has nothing to do with the velocity of the train. This also indicates the following conclusion:

\emph{\textbf{Corollary 1:}} \emph{Under the same assumptions of Proposition 1, a train travelling from point $-l$ to point $l$ receives the same service ratio of a base station no matter how fast it travels at a constant velocity.}

\begin{proof} The service that a train can receive from point$-l$ to point $l$ is $\int_{-l/v}^{l/v}\log(1+\frac{\gamma}{(d_0^2+(v\tau)^2)^{\frac{\alpha}{2}}})\,\mathrm{d}\tau $, correspondingly, the service ratio $\eta_l$ is
\begin{equation}\label{equ_Clsim}
  \eta_l=\frac{\int_{0}^{l}\log(1+\frac{\gamma}{(d_0^2+x^2)^{\frac{\alpha}{2}}})\,\mathrm{d}x }{\int_{0}^{\infty}\log(1+\frac{\gamma}{(d_0^2+x^2)^{\frac{\alpha}{2}}})\,\mathrm{d}x }.
\end{equation}
Since the denominator is a constant, the nominator depends only on $l$, so $\eta_{l}$ is a constant determined by $l$, this concludes our proof.
\end{proof}

For $\alpha=2$, Eqn. (\ref{equ_etavsd}) can be expressed as:
\begin{equation}\label{equ_etadis2}
  \eta=\frac{\frac{d_s}{2}\ln(1+\frac{\gamma}{B})+2A\arctan{\frac{\frac{d_s}{2}}{A}}-2d_0\arctan{\frac{\frac{d_s}{2}}{d_0}}}{\pi(A-d_0)},
\end{equation}
where $A=\sqrt{\gamma+d_0^2}$, $B=d_0^2+(\frac{d_s}{2})^2$. For $\alpha>2$, the indefinite integral cannot be expressed in an explicit form of simple functions.

\begin{figure}
\centering
\includegraphics[width=0.48\textwidth]{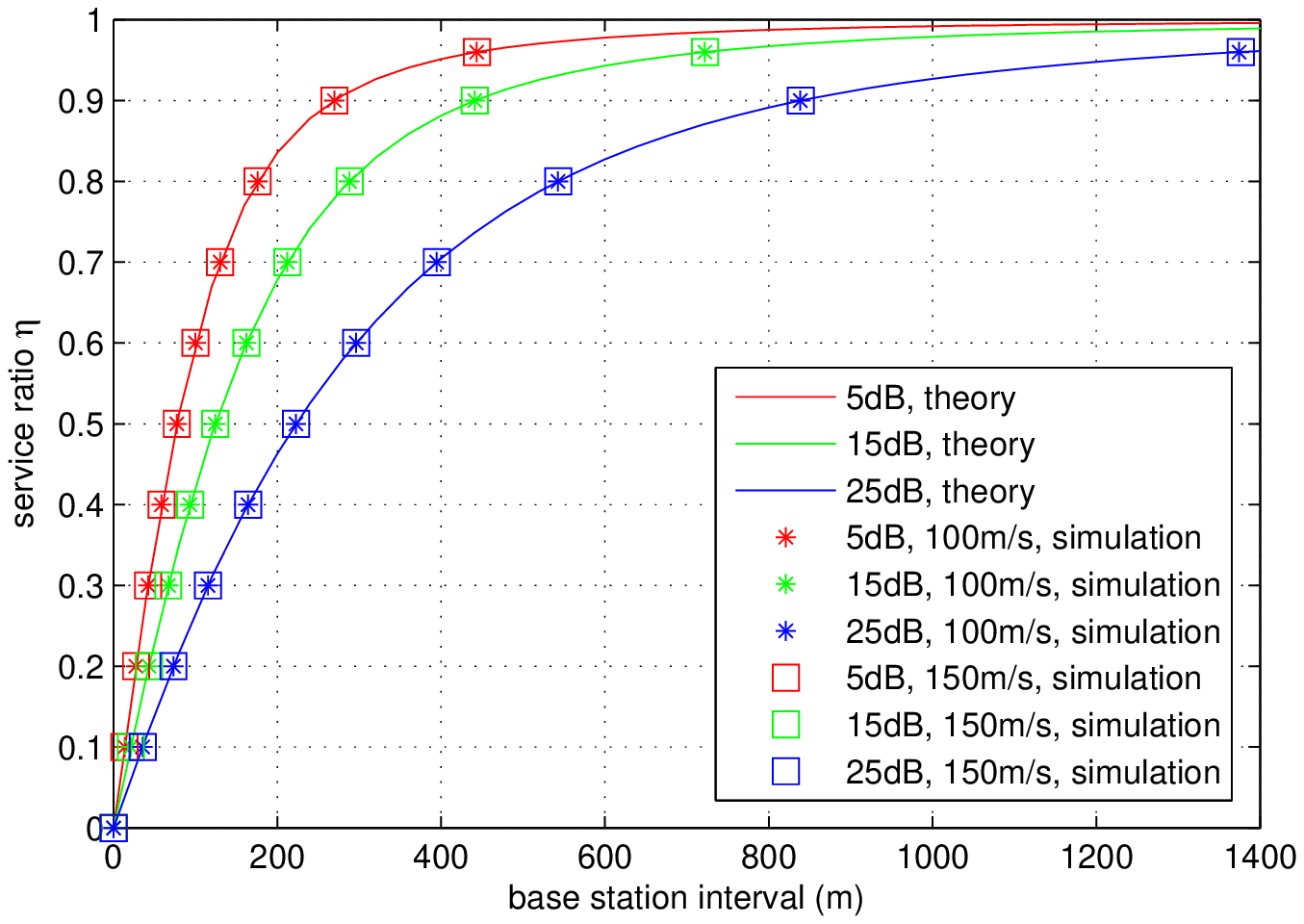}
\caption{Service ratio versus base station interval for $SNR=5,15,25 dB$, $v=100m/s,150m/s$, $d_0=50m$, $\alpha=3$.}\label{fig_etaint}
\end{figure}

The service ratio versus base station interval curves are shown in Fig. \ref{fig_etaint}. In this figure, theoretical curves are plotted in solid lines while simulation results are drawn with points. It can be seen that the simulation results match well with theoretical results, and the results with different velocities coincide when the $SNR_0$s are the same, which demonstrates the correctness of Proposition 1. Besides, it can be seen from this figure that for the same service distance $d_s$, smaller transmitter SNR actually gets higher service ratio $\eta$.

Next, we show that choice of base station interval using the service ratio criterion is equivalent to using the minimum channel capacity criterion. It is easy to see from the above discussions that the mapping from service ratio to base station interval is one to one. By substituting Eqn. (\ref{equ_dssimp}) into
\begin{equation}
R_m=\log(1+\frac{\gamma}{(d_{0}^{2}+(\frac{d_{s}}{2})^{2})^{\frac{\alpha}{2}}}),
\end{equation}
where $R_m$ is the minimum channel capacity, we get
\begin{equation}\label{}
  R_m=\log(1+\frac{\gamma}{(d_{0}^{2}+(\frac{f_1^{-1}(C_1\eta)}{2})^{2})^{\frac{\alpha}{2}}}),
\end{equation}
The equation above shows that the mapping from service ratio to minimum rate is also one to one, thus for the minimum rate, we can always find a corresponding service ratio to get the same base station interval. In fact, when evaluating what percentage should be chosen, the minimum channel capacity may provide a tractable guidance since it is equivalent to the service ratio criterion in determining the interval of base stations.

\subsection{Choice of Base Station Interval given a Service Amount Requirement}\label{sec_bsisa}

It was shown in Section \ref{sec_bsisr} that given a service ratio requirement, the base station interval did not depend on the velocity of the train. In this section, we consider another criterion where base stations are required to provide a certain amount of service to the passing train. Suppose that through coordination and prediction, a base station has already buffered the required data for the train, and needs to finish the transmission before the train leaves its service region. Then, we show that if we use the service amount to determine the service distance, the service distance or base station interval will increase with the velocity of the train. This is clearly shown in the following proposition.

\emph{\textbf{Proposition 2:}}  \emph{If the transmitter SNR, the path loss exponent and the distance between the base station and the rail are fixed, the base station interval decided by a given service amount increases with the velocity of the train.}

\begin{proof}
Assuming that the base station interval corresponding to a given service amount $\mathbb{S}$ is $d_s$, then
\begin{eqnarray}\label{equ_servspe}
\mathbb{S}=\frac{2}{v}\int_{0}^{\frac{d_s}{2}}\log(1+\frac{\gamma}{(d_0^2+x^2)^{\frac{\alpha}{2}}})\,\mathrm{d}x.
\end{eqnarray}
Rewrite this equation, we have
\begin{equation}\label{equ_serint}
  \int_{0}^{\frac{d_s}{2}}\log(1+\frac{\gamma}{(d_0^2+x^2)^{\frac{\alpha}{2}}})\,\mathrm{d}x=\frac{v\mathbb{S}}{2}.
\end{equation}
The left side of the above equation is the same with Eqn. (\ref{equ_Cetasim}), which we denote as $f_1(d_s)$, since it is a monotonically increasing function of $d_s$ and thus invertible, we can get
\begin{equation}
  d_s=f_1^{-1}(\frac{v\mathbb{S}}{2}).
\end{equation}
So, the base station interval $d_s$ is monotonically increasing with the velocity of the train $v$ since $f_1^{-1}(\frac{v\mathbb{S}}{2})$ is also a monotonically increasing function.
\end{proof}

For the case $\alpha=2$, there is a closed form for the integral of Eqn. (\ref{equ_serint}), and the velocity of the train can be given as
\begin{equation}\label{equ_clsspdis}
  v=\frac{2}{\mathbb{S}\ln2}[\frac{d_s}{2}\ln(1+\frac{\gamma}{B})+2A\arctan{\frac{\frac{d_s}{2}}{A}}
  -2d_0\arctan{\frac{\frac{d_s}{2}}{d_0}}].
\end{equation}
where $A=\sqrt{\gamma+d_0^2}$, $B=d_0^2+(\frac{d_s}{2})^2$. For $\alpha>2$, there is no explicit form for the integral.
\begin{figure}
  \centering
  \includegraphics[width=0.48\textwidth]{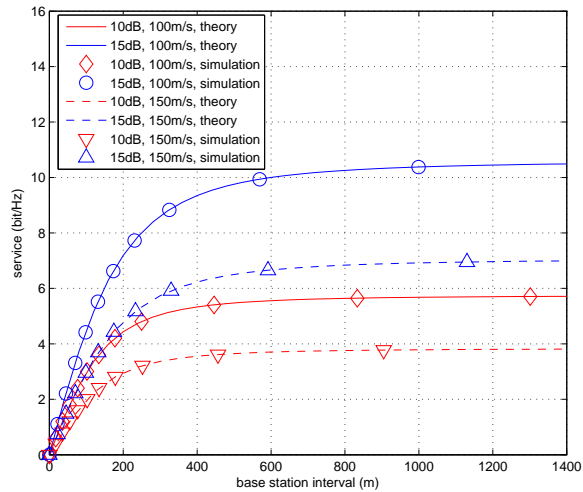}\\
  \caption{service versus base station interval for $SNR=10,15 dB$, $v=100m/s,150m/s$, $d_0=50m$, $\alpha=3$.}\label{fig_serint}
\end{figure}

The relations between service and base station interval at different $SNR_0$s and velocities are shown in Fig. \ref{fig_serint}. Again, simulation results agree well with theoretical analysis, and comparing the solid line with the dashed line, we can see that for the same service, the dashed line with higher velocity requires longer base station interval, which is a direct proof of Proposition 2.

\section{Discussion on the Arc Rail}\label{sec_arcrail}
\begin{figure}
  \centering
  \includegraphics[width=0.48\textwidth]{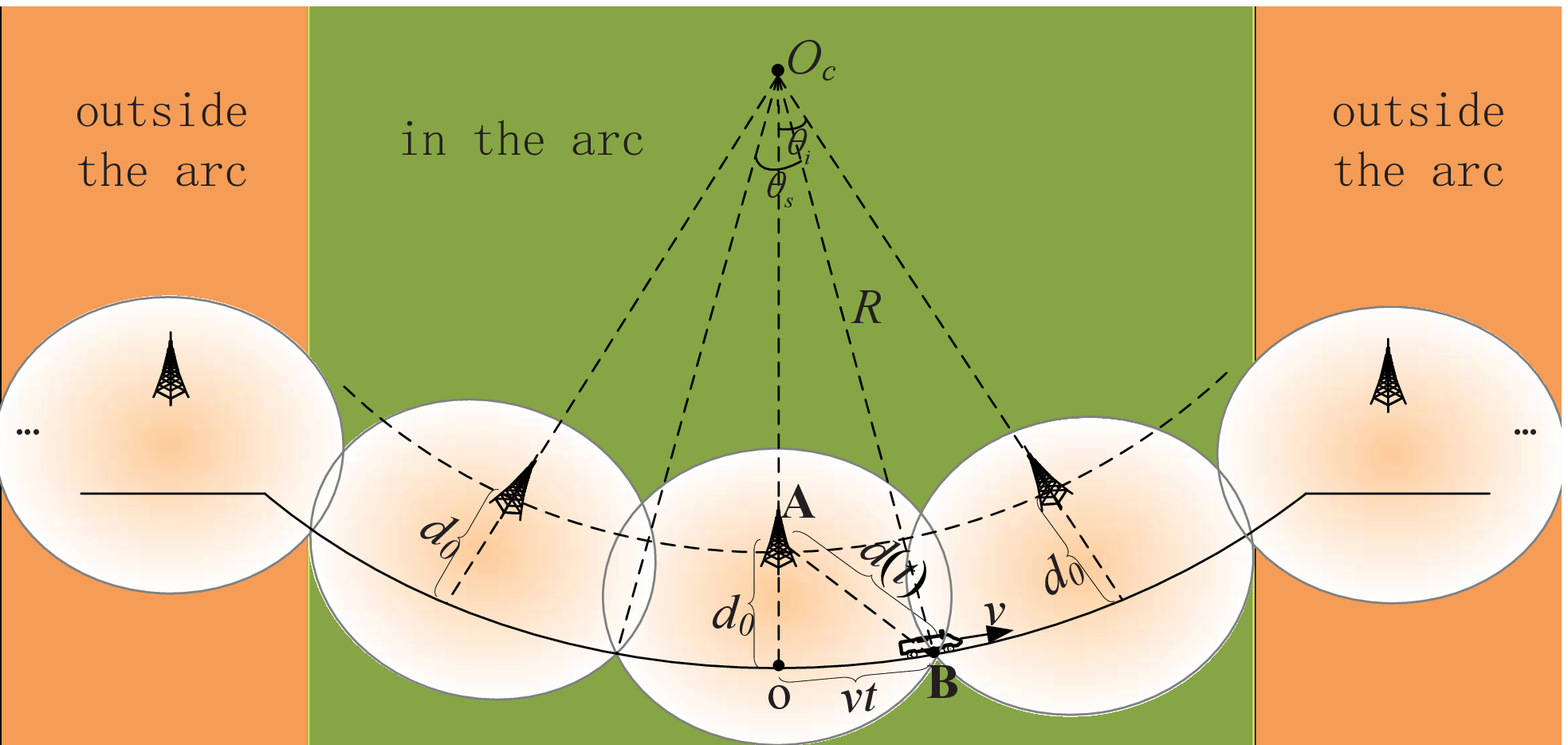}\\
  \caption{Base station coverage along an arc rail.}\label{fig_arcrail}
\end{figure}

\subsection{Service of a Base Station}
In practical environments, the railway may not always be straight due to various geological conditions, it may bend to avoid lakes, mountains or high buildings  and forms an arc. In this section, we consider base station deployment along an arc rail, as shown in Fig. \ref{fig_arcrail}. The distance from the base station to the rail is $d_0$, the radius of the arc is $R$, so the distance between the center of the arc and the base station is $R_s=R-d_0$. Assuming that the train is moving at a constant velocity $v$, the moment when the train passes the point $o$ is time $0$, then at time $t$, the distance between base station A and the train is:
\begin{equation}\label{equ_distarc}
  d(t)=\sqrt{R^2+R_{s}^{2}-2RR_s\cos(\frac{vt}{R})}.
\end{equation}
Correspondingly, the channel capacity at time $t$ is:
\begin{equation}\label{equ_chcaparc}
  C(t)=\log(1+\frac{\gamma}{(R^2+R_s^2-2RR_s\cos(\frac{vt}{R}))^{\frac{\alpha}{2}}}).
\end{equation}
Assuming that the train is within the arc from time $t_1$ to $t_2$, then the service of this time period is
\begin{equation}\label{equ_sequaarc}
  S(t_1,t_2)=\int_{t_1}^{t_2}\log(1+\frac{\gamma}{(R^2+R_s^2-2RR_s\cos(\frac{v\tau}{R}))^{\frac{\alpha}{2}}})\,\mathrm{d}\tau.
\end{equation}
The integration over time can also be changed to be the integration over distance
\begin{equation}\label{equ_sequaarcd}
  S(x_1,x_2)=\frac{1}{v}\int_{x_1}^{x_2}\log(1+\frac{\gamma}{(R^2+R_s^2-2RR_s\cos(\frac{l}{R}))^{\frac{\alpha}{2}}})\,\mathrm{d}l,
\end{equation}
where $x_1=vt_1$, $x_2=vt_2$.

We assume that the radius of the arc $R$ is much larger than the distance between the base station and the railway $d_0$, this is reasonable since in many cases $R$ can be several or tens of kilometers while $d_0$ may be only tens of or several hundred meters. For instance, along the Beijing-Shanghai high-speed railway, the radius of the arc from Nanjing to Changzhou is approximately 50km, while the distance between the BS and the railway may be only 100m.  Based on this assumption, for the base stations in the arc, their total service is ``in the arc", since the channel capacity of the base station outside the arc tends to zero and thus can be ignored. For the base stations at the edge of the arc, the railway is an irregular curve which we delay our discussion to the next section. The total service of a base station in the arc can be expressed as

\begin{equation}\label{equ_tswa}
  S_{tot}=\frac{1}{v}\int_{-x_{\infty}}^{x_{\infty}}\log(1+\frac{\gamma}{(R^2+R_s^2-2RR_s\cos(\frac{l}{R}))^{\frac{\alpha}{2}}})\,\mathrm{d}l,
\end{equation}
where $x_{\infty}$ is a point ``in the arc" and at the same time large enough so that the service outside $[-x_{\infty},x_{\infty}]$ can be ignored.

It is obvious that the total service is inversely proportional to the velocity of the train, this is the same with the line rail. In fact, the conclusions regarding the base station arrangement issue in the previous section apply to this case. Different from the line rail, we use the central angle of the arc to represent the``interval" of base stations since the position of each base station in this arc can be solely decided by this angle. Besides, we also assume that the service angle $\theta_s$ of all base stations are the same, in this case, $\theta_i=\theta_s$, where $\theta_i$ is the angle between two neighboring base stations or the base station``interval", so we'll not distinguish them later and use $\theta_i$ for both of them.

\subsection{Choice of Service Angle given a Service Ratio Requirement}
We have the following proposition when deciding the base station``interval" using the service ratio.

\textbf{\emph{Proposition 3:}} \emph{If the base station is required to provide a certain service ratio in its service region, the central angle between two base stations or the base station``interval" is irrelevant to the velocity of the train when the radius of the arc $R$, the distance between the base station and the railway $d_0$, the path loss exponent $\alpha$ and the transmitter SNR $\gamma$ are given.}

\begin{proof} If from time $-t_s$ to $t_s$, the train is in the service region of base station A as in Fig. \ref{fig_arcrail}, then the service ratio is
\begin{equation}\label{equ_drwar}
  \eta=
  \frac{\int_{-t_s}^{t_s}\log(1+\frac{\gamma}{(R^2+R_s^2-2RR_s\cos(\frac{v\tau}{R}))^{\frac{\alpha}{2}}})\,\mathrm{d}\tau}
  {\frac{1}{v}\int_{-x_{\infty}}^{x_{\infty}}\log(1+\frac{\gamma}{(R^2+R_s^2-2RR_s\cos(\frac{l}{R}))^{\frac{\alpha}{2}}})\,\mathrm{d}l}.
\end{equation}
Note that $\theta_i R=2vt_s$, the above equation then becomes
\begin{equation}\label{equ_drwarl}
  \eta=
  \frac{\int_{-\frac{\theta_i R}{2}}^{\frac{\theta_i R}{2}}\log(1+\frac{\gamma}{(R^2+R_s^2-2RR_s\cos(\frac{l}{R}))^{\frac{\alpha}{2}}})\,\mathrm{d}l}
  {\int_{-x_{\infty}}^{x_{\infty}}\log(1+\frac{\gamma}{(R^2+R_s^2-2RR_s\cos(\frac{l}{R}))^{\frac{\alpha}{2}}})\,\mathrm{d}l}.
\end{equation}
By rewriting it, we have
\begin{multline}\label{equ_drthetaar}
  \int_{0}^{\frac{\theta_i R}{2}}\log(1+\frac{\gamma}{(R^2+R_s^2-2RR_s\cos(\frac{l}{R}))^{\frac{\alpha}{2}}})\,\mathrm{d}l \\
  =\eta\int_{0}^{x_{\infty}}\log(1+\frac{\gamma}{(R^2+R_s^2-2RR_s\cos(\frac{l}{R}))^{\frac{\alpha}{2}}})\,\mathrm{d}l.
\end{multline}
The left side of the above equation is a monotonically increasing function of $\theta_i$ if $R$, $d_0$, $\alpha$, $\gamma$ are given, we denote as $f_2(\theta_i)$, the definite integration on the right side is a constant, which we represent as $C_2$, then the above equation can be written in a simple form
\begin{equation}
  f_2(\theta_i)=C_2\eta.
\end{equation}
Since $f_2$ is invertible, we have
\begin{equation}\label{equ_arcinra}
\theta_i=f_2^{-1}(C_2\eta)
\end{equation}
As can be seen from the above equation, the service angle $\theta_i$ is a monotonically increasing function of $\eta$, if $\eta$ is given, it is a constant regardless of the velocity of the train.
\end{proof}

\subsection{Choice of Service Angle given a Service Amount Requirement}
When using the service amount (the service amount is less than the total service) to decide the service angle, we have

\textbf{\emph{Proposition 4:}} \emph{The service angle decided by a certain service amount increases with the velocity of the train.}
\begin{proof}
If the base station is required to provide a certain service amount, denoted as $\mathbb{S}$, to the train from $-t_s$ to $t_s$, then
\begin{equation}\label{equ_seranwac}
 \mathbb{S}=\frac{2}{v}\int_{0}^{\frac{\theta_i R}{2}}\log(1+\frac{\gamma}{(R^2+R_s^2-2RR_s\cos(\frac{l}{R}))^{\frac{\alpha}{2}}})\,\mathrm{d}l.
\end{equation}
Similarly, the above equation can be expressed in a simple form as in the proof of Proposition 3, which is
\begin{equation}
\theta_i=f_2^{-1}(\frac{\mathbb{S}v}{2})
\end{equation}
Since $f_2$ is a monotonically increasing function, its inverse is also monotonically increasing, so, the service angle $\theta_i$ will increase with the velocity of the train.
\end{proof}

The line rail and the arc rail are two typical scenarios when we deploy the base stations, however, they are only special cases in the actual environments. Next, we'll discuss base station arrangement for the general curve rail. In fact, it will be shown that the same conclusions can be applied for any kind of curves of the railway, as long as the velocity of the train is a constant.

\section{Discussion on the Irregular Curve Rail}\label{sec_anycurverail}

\subsection{Service of a Base Station}
In this section, we consider a more general case when the railway is an irregular curve. We will show that the results for the line rail and the arc rail hold for this scenario provided that the train is travelling at a constant velocity.

\begin{figure}
  \centering
  \includegraphics[width=0.48\textwidth]{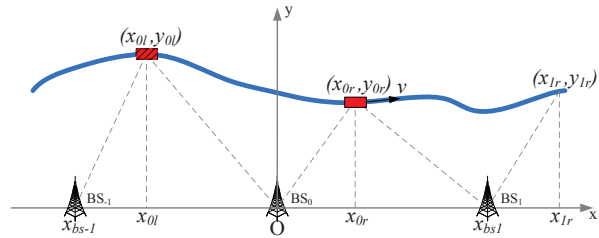}\\
  \caption{Base station arrangement along an irregular curve rail.}\label{fig_anycur}
\end{figure}

The model of an irregular curve rail is shown in Fig. \ref{fig_anycur}. Different from previous sections, we establish a coordinate setting a base station as the original point in this figure. To prove the same conclusions for this kind of rail, we firstly introduce the following lemma:

\emph{\textbf{Lemma 1:} The service amount that a constant-velocity travelling train receives between any two points of the rail is decided by the transmitter SNR, the path loss exponent, the curve of the rail and the velocity of the train. In particular, it is inversely proportional to the velocity of the train.}
\begin{proof}
We formulate the curve of the rail as a function $y=f(x)$ in the coordinate system. Since the rail is continuous and smooth, the function is continuous and differentiable. If the train passes points $(x_1,y_1)$ and $(x_2,y_2)$ at $t_1$ and $t_2$ respectively, then at time $t\in(t_1,t_2)$, the position of the train $(x_t,y_t)$ satisfies
\begin{equation}\label{equ_posac}
  v*(t-t_1)=\int_{x_1}^{x_t}\sqrt{1+(f'(x))^2}\,\mathrm{d}x.
\end{equation}
Denote $\int_{x_1}^{x_t}\sqrt{1+(f'(x))^2}\,\mathrm{d}x$ as $\varphi(x_t)$, thus \begin{equation}\label{equ_tposac}
  t=\frac{1}{v}\varphi(x_t)+t_1.
\end{equation}
It can be seen that $\varphi$ is monotonically increasing with $x_t$, therefore, the projection from $x_t$ to $t$ is one to one, and we can get the inverse projection as
\begin{equation}\label{equ_postac}
  x_t=\varphi^{-1}(v*(t-t_1)).
\end{equation}
The service from $t_1$ to $t_2$ is
\begin{equation}\label{equ_serquac}
  S(t_1,t_2)=\int_{t_1}^{t_2}\log(1+\frac{\gamma}{(x_{t}^{2}+y_{t}^{2})^{\frac{\alpha}{2}}})\,\mathrm{d}t.
\end{equation}
Note that $y_t=f(x_t)$, and use the substitution $t=\frac{1}{v}\varphi(x)+t_1.$, Eqn. (\ref{equ_serquac}) becomes:
\begin{equation}\label{equ_serqudist}
  S(t_1,t_2)=\frac{1}{v}\int_{x_1}^{x_2}\log(1+\frac{\gamma}{(x^2+f^{2}(x))^{\frac{\alpha}{2}}})\varphi'(x)\,\mathrm{d}x.
\end{equation}

The above equation indicates that the service that the train can receive between $(x_1,y_1)$ and $(x_2,y_2)$ is determined by the transmitter SNR, the path loss exponent, the curve of the rail which influences the value of $f(x)$ and $\varphi(x)$, and the velocity of the train. In particular, given the transmitter SNR, the path loss exponent and the curve, it is inversely proportional to the velocity of the train.
\end{proof}

Since the service between any two points is inversely proportional to the velocity of the train, the following corollary can be derived.

\textbf{\emph{Corollary 2:}} \emph{The service that a train receives between any two points accounts for the same ratio of the total service of a base station as long as the train is moving at a constant velocity.}

\subsection{The Line along which to Deploy Base Stations}

Like in the above two cases, we should normally deploy base stations along a curve line which is parallel to the railway, but we find that although the service requirements can be formulated easily, the calculation of the position of each base station is quite difficult. To simplify this problem, we take an approximation approach to arrange the base stations along a straight line which is ``parallel" to the curve. This approach works well when the variation of the curve is not large, and for curves with large variation, we can divide them into segments with smaller variations since the railway usually doesn't turn sharply within a short distance, but rather remains smooth over a long distance. So in later discussions, we always assume the variation of the curve is not large.

To decide the line along which to deploy base stations, we first establish a coordinate arbitrarily provided that the railway is always on the upper side of x axis. Then the position of each point on the railway can be determined by measuring the distances from the point to any other two points on x axis. Select some points which can represent the curve of the rail approximately (we can use uniform sampling for convenience), and with these points we determine a line using the method of least squares. For instance, if the points $(x_i,y_i),i=1,2,\cdots,n$ on the railway are chosen, the line decided by the method of least squares is \[y=ax+b,\] with
\begin{equation}\label{equ_slp}
  a=\frac{\sum_{i=1}^{n}{(x_i-\overline{x})(y_i-\overline{y})}}{\sum_{i=1}^{n}{(x_i-\overline{x})^2}},
\end{equation}
\begin{equation}\label{equ_intc}
  b=\overline{y}-a*\overline{x},
\end{equation}
where $\overline{x}=\frac{1}{n}\sum_{i=1}^{n}{x_i}$, $\overline{y}=\frac{1}{n}\sum_{i=1}^{n}{y_i}$.

\begin{figure}
  \centering
  \includegraphics[width=0.48\textwidth]{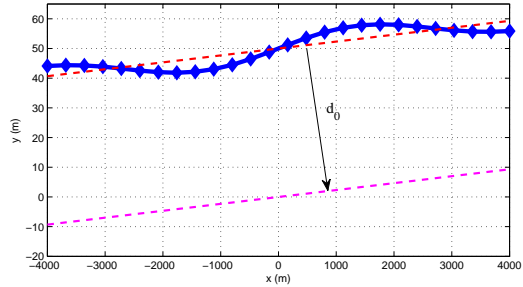}\\
  \caption{The method of least squares to determine the line.}\label{fig_leastsqr}
\end{figure}

The line to deploy base stations is translated $d_0$ distance away from this least-squares-determined line. So its expression is
\begin{equation}\label{equ_linebs}
  y=a*x+b-d_0*\sqrt{1+a^2}.
\end{equation}
The realization of this method is shown in Fig. \ref{fig_leastsqr}.

For the convenience of later discussions, we use the deployment line as the x axis and choose one point arbitrarily in the line as the original point. The curve of the rail in this coordinate system is denoted as $f(x), x\in[x_l,x_r]$. Besides, as in previous sections, we choose the service region of each base station symmetrically around it.

%

\subsection{Algorithms to Deploy Base Stations given Service Requirements }
We then show how to use service ratio to arrange base stations. Assuming that each base station is required to provide a given service ratio $\eta$, then for the first base station $BS_0$ with service region from $x_{0l}$ to $x_{0r}$ ($x_{0l}=-x_{0r}$), we have
\begin{eqnarray}\label{equ_serdisac}
  \eta &=& \frac{\frac{1}{v}\int_{x_{0l}}^{x_{0r}}\log(1+\frac{\gamma}{(x^2+f^{2}(x))^{\frac{\alpha}{2}}})\varphi'(x)\,\mathrm{d}x}
  {\frac{1}{v}\int_{-\infty}^{\infty}\log(1+\frac{\gamma}{(x^2+f^{2}(x))^{\frac{\alpha}{2}}})\varphi'(x)\,\mathrm{d}x}
  \nonumber \\
  &=&
  \frac{\int_{x_{0l}}^{x_{0r}}\log(1+\frac{\gamma}{(x^2+f^{2}(x))^{\frac{\alpha}{2}}})\varphi'(x)\,\mathrm{d}x}
  {\int_{-\infty}^{\infty}\log(1+\frac{\gamma}{(x^2+f^{2}(x))^{\frac{\alpha}{2}}})\varphi'(x)\,\mathrm{d}x}.
\end{eqnarray}

From Eqn. (\ref{equ_serdisac}), we can get the value of $x_{0r}$ uniquely since $\eta$ is monotonically increasing with $x_{0r}$. Since there is no explicit expression of $x_{0r}$ as a function of $\eta$, we can calculate the value of $x_{0r}$ using dichotomy.

Let the position of $BS_k$ be $x_{bs_k}$, in order to provide the given service ratio, for $BS_k$ with service region $[x_{kl},x_{kr}]$, we have
\begin{eqnarray}\label{equ_serdis2ac}
  \eta &=& \frac{\frac{1}{v}\int_{x_{kl}}^{x_{kr}}\log(1+\frac{\gamma}{((x-x_{bs_k})^2+f^{2}(x))^{\frac{\alpha}{2}}})\varphi'(x)\,\mathrm{d}x}
  {\frac{1}{v}\int_{-\infty}^{\infty}\log(1+\frac{\gamma}{((x-x_{bs_k})^2+f^{2}(x))^{\frac{\alpha}{2}}})\varphi'(x)\,\mathrm{d}x}
  \nonumber \\
  &=&
  \frac{\int_{x_{kl}}^{x_{kr}}\log(1+\frac{\gamma}{((x-x_{bs_k})^2+f^{2}(x))^{\frac{\alpha}{2}}})\varphi'(x)\,\mathrm{d}x}
  {\int_{-\infty}^{\infty}\log(1+\frac{\gamma}{((x-x_{bs_k})^2+f^{2}(x))^{\frac{\alpha}{2}}})\varphi'(x)\,\mathrm{d}x},
\end{eqnarray}
where $x_{kr}=2x_{bs_k}-x_{kl}$. Here $\eta$ may not be strictly monotonically increasing with $x_{bs_k}$, it may decrease at some values of $x_{bs_k}$, so there could be numerous values of $x_{bs_k}$ satisfying Eqn. (\ref{equ_serdis2ac}), in this case, we choose the smallest value for $k>0$ and largest value for $k<0$ as $x_{bs_k}$. The existence of the value of $x_{bs_k}$ is shown in the following lemma.

\emph{\textbf{Lemma 2}} \emph{Given $x_{kl}$, there exists at least one $x_{bs_k}$ satisfying Eqn. (\ref{equ_serdis2ac}).}
\begin{proof}
Substituting $x_{kr}=2x_{bs_k}-x_{kl}$ in Eqn. (\ref{equ_serdis2ac}), we have
\begin{equation}\label{equ_etabspos}
  \eta=\frac{\int_{x_{kl}}^{2x_{bs_k}-x_{kl}}\log(1+\frac{\gamma}{((x-x_{bs_k})^2+f^{2}(x))^{\frac{\alpha}{2}}})\varphi'(x)\,\mathrm{d}x}
  {\int_{-\infty}^{\infty}\log(1+\frac{\gamma}{((x-x_{bs_k})^2+f^{2}(x))^{\frac{\alpha}{2}}})\varphi'(x)\,\mathrm{d}x}.
\end{equation}
It is easy to see that $\eta$ is continuous with $x_{bs_k}$ since both the denominator and the nominator are continuous with $x_{bs_k}$ and the denominator is not equal to zero. For $x_{bs_k}=x_{kl}$, $\eta=0$, and for $x_{bs_k} \rightarrow \infty$, $\eta \rightarrow 1$. This can be shown in the following equation,
\begin{equation}\label{equ_etabspos_s}
  \eta=\frac{\int_{x_{kl}-x_{bs_k}}^{x_{bs_k}-x_{kl}}\log(1+\frac{\gamma}{(x^2+f^{2}(x+x_{bs_k}))^{\frac{\alpha}{2}}})\varphi'(x+x_{bs_k})\,\mathrm{d}x}
  {\int_{-\infty}^{\infty}\log(1+\frac{\gamma}{(x^2+f^{2}(x+x_{bs_k}))^{\frac{\alpha}{2}}})\varphi'(x+x_{bs_k})\,\mathrm{d}x},
\end{equation}
with $x_{bs_k}$ approaching $\infty$, $x_{kl}-x_{bs_k}$ and $x_{bs_k}-x_{kl}$ tend to $-\infty$, $\infty$ respectively, thus $\eta$ tends to $1$. So for a $\eta \in [0,1]$, there exists at least one $x_{bs_k}$ satisfying Eqn. (\ref{equ_serdis2ac}).
\end{proof}
Again, there is no explicit expression of $x_{bs_k}$ as a function of $\eta$, so the value of $x_{bs_k}$ should be calculated using some numerical methods such as dichotomy. The position of each base station can be determined successively. To summarize the above approach, we have the following algorithm \\

\begin{tabular}{l @{ } l}
  \hline
  \multicolumn{2}{c}{\textbf{Algorithm 1:} BS Deployment given Service Ratio} \\ \multicolumn{2}{c}{Requirement} \\
  \hline
  1 & \textbf{Initialization:} $x_{bs_0}\leftarrow 0$, calculate $x_{0r}$ from Eqn. (\ref{equ_serdisac}); \\
  2 & \textbf{for} $k=1,2,\cdots$ \textbf{do} \\
   &  $x_{kl}\leftarrow x_{(k-1)r}$, calculate $x_{bs_k}$ and\\
   & $x_{kr}=2*x_{bs_k}-x_{kl}$ from Eqn. (\ref{equ_serdis2ac});  \\
   & \textbf{end}  \\
  3 & \textbf{for} $k=-1,-2,\cdots$ \textbf{do} \\
   &  $x_{kr}\leftarrow x_{(k+1)l}$, calculate $x_{bs_k}$ and \\
   & $x_{kl}=2*x_{bs_k}-x_{kr}$ from Eqn. (\ref{equ_serdis2ac});  \\
   & \textbf{end}  \\
  \hline
\end{tabular}
\\

As can be seen from the above algorithm, the positions of base stations are irrelevant to the velocity of the train. That is, the velocity of the train doesn't affect the arrangement of base stations.

As in previous sections, we can also use the service amount to decide the position of each base station. The line to deploy base stations is determined using the same method as in Fig. \ref{fig_leastsqr}. Different from the service ratio requirement, each base station is demanded a certain amount of service, denoted by $\mathbb{S}$. The similar algorithm to determine the position of each base station is:
For the base station $BS_0$ with service region from $x_{0l}$ to $x_{0r}$ ($x_{0l}=-x_{0r}$), we have
\begin{equation}\label{equ_serquadac}
  \mathbb{S}=\frac{1}{v}\int_{x_{0l}}^{x_{0r}}\log(1+\frac{\gamma}{(x^2+f^{2}(x))^{\frac{\alpha}{2}}})\varphi'(x)\,\mathrm{d}x.
\end{equation}
The values of $x_{0l}$ and $x_{0r}$ can be uniquely calculated.

Let the position of $BS_k$ be $x_{bs_k}$, in order to provide the given amount of service, for $BS_k$, we have
\begin{equation}\label{equ_serquad2ac}
  \mathbb{S}=\frac{1}{v}\int_{x_{kl}}^{x_{kr}}\log(1+\frac{\gamma}{((x-x_{bs_k})^2+f^{2}(x))^{\frac{\alpha}{2}}})\varphi'(x)\,\mathrm{d}x,
\end{equation}
where $x_{kr}=2x_{bs_k}-x_{kl}$.
From Eqn. (\ref{equ_serquad2ac}), we can calculate $x_{bs_k}$ numerically. Again, the value of $x_{bs_k}$ may be multiple, and we choose the smallest value for $k>0$ and largest value for $k<0$ as the position of $BS_k$.  The position of each base station can be determined successively. The approach is given in Algorithm 2. \\

\begin{tabular}{ l @{  } l }
  \hline
   \multicolumn{2}{c}{\textbf{Algorithm 2:} BS Deployment given Service Amount}
    \\ \multicolumn{2}{c}{Requirement} \\
   \hline
  1 & \textbf{Initialization:} $x_{bs_0}\leftarrow 0$, calculate $x_{0r}$ from Eqn. (\ref{equ_serquadac}); \\
  2 & \textbf{for} $k=1,2,\cdots$ \textbf{do} \\
    & $x_{kl}\leftarrow x_{(k-1)r}$, calculate $x_{bs_k}$ and\\
    & $x_{kr}=2*x_{bs_k}-x_{kl}$ from Eqn. (\ref{equ_serquad2ac});  \\
    & \textbf{end}  \\
  3 & \textbf{for} $k=-1,-2,\cdots$ \textbf{do} \\
    & $x_{kr}\leftarrow x_{(k+1)l}$, calculate $x_{bs_k}$ and \\
    & $x_{kl}=2*x_{bs_k}-x_{kr}$ from Eqn. (\ref{equ_serquad2ac});  \\
   & \textbf{end}  \\
  \hline
\end{tabular}
\\

In this case, the velocity of the train influences the choice of the position of each base station. The higher the velocity of the train, the longer the service distance of each base station and the larger the interval of adjacent base stations.

 \textit{Example} Let us assume a length of irregular curve of the railway can be shaped by a function $f(x)=2\sin(4\pi*10^{-4}*x)+3\sin(4\pi*10^{-5}*x)+d_0$. The parameter settings are as follows: the average distance from base stations to the railway is $d_0=50m$, the path loss exponent is $\alpha=3$, the velocity of the train is $v=100m/s$, the received SNR at point O is $SNR_0=25dB$.  Then we present 10 boundary points using each of the proposed two algorithms: Algorithm 1 and Algorithm 2. The results are listed in TABLE \ref{tbl_boundary}, where the positions of base stations lie in the middle of two adjacent boundary points. It is hard to see that the interval between two neighboring base stations is not equal but with almost the  same value. The difference  is due to the fluctuation of the railway.

\begin{table}
\begin{center}
\caption{Boundary points calculated by using Algorithm 1 and Algorithm 2 } \label{tbl_boundary}
\begin{tabular}{|c|c|c|c|c|c|c|c|c|c|c|} \hline
\backslashbox{Algorithms}{Boundary points}
&$x_{-5r}$&$x_{-4r}$&$x_{-3r}$&$x_{-2r}$&$x_{-1r}$&$x_{0r}$&$x_{1r}$&$x_{2r}$&$x_{3r}$&$x_{4r}$ \\\hline
  Algorithm 1 $\eta=0.8$ &-2781&-2273&-1523&-773&-262& 262&777&1324&1871&2387 \\
  \hline
  Algorithm 2 $\mathbb{S}=23$ &-2434&-1891&-1352&-814&-271&271&816	&1360&1904&2448\\
  \hline
\end{tabular}
\end{center}
\end{table}

\section{Discussion on the On-Off Transmission Strategy of Base Stations}\label{sec_scrttr}

In previous sections, we focused on the decision of base station intervals. And we see that the service provided by a base station is influenced by the velocity of the train, i.e., the higher the velocity, the fewer the service. To finish the required transmission for trains with different velocities, some kind of adaptation techniques may be necessary for the base station, for instance, adaptive power allocation, or adaptive service region management. In this section, we introduce an on-off transmission strategy as an application of our conclusions to solve this problem. Specifically, the transmission problem is described as follows:

A base station buffers a certain amount of data, which is equal to its total service. It is required to transmit a given ratio or amount of the buffer data to a constant-velocity-travelling train in its service region. With constant transmission power, when should it start and stop its transmission?

Using the results in previous sections, one can get the following answers: To transmit a certain ratio of the data, the distance the train travels during the transmission process has nothing to do with the velocity of the train, which can be obtained by solving $d_s$ from Eqn. (\ref{equ_Cetasim}). This is true since both the amount of data the train receives and the maximum amount of data the base station buffers decrease with $v$. However, to transmit a certain amount of data, the velocity of the train does affect the transmission's beginning and ending time, their relation is shown in Eqn. (\ref{equ_serint}). Intuitively, the higher the velocity, the longer the distance the station needs to serve, and the earlier (later) it needs to start (stop) the transmission.

\section{Conclusion}\label{sec_conclusion}

In this work, we discussed the base station deployment issue from the perspective of service. Specifically, we analyzed how to deploy base stations along a line rail, an arc rail and an irregular curve rail. For each of the three different cases, we discussed the relationship between the base station interval and the velocity of the train in two scenarios. In the first scenario, we guaranteed that a certain ratio of the station's total service be provided, and proved that the interval of adjacent base stations is irrelevant to the velocity of the train. In the second scenario, if a given amount of service is required, the interval will increase with the velocity of the train. In addition, we applied these results in the analysis of the on-off transmission strategy of base stations.

It needs to be noted that the analysis in this paper has some limitations: Firstly, the model of the channel is a much simplified model since we didn't consider the effect of fading. Actually, when fading is added in the expression of channel capacity, the conclusions may have some changes since the capacity would be a random process rather than a definite function of time, so further study is needed when fading is considered, and this may be one of the directions of our future work. Secondly, the power of a base station is assumed to be constant in this paper, however, some adaptive power allocation method may improve the performance greatly. Thirdly, we only discussed the decision of base station interval based on the assumption that other design parameters are given, however, parameters like the distance between the base station and the railway or the height of the transmitter can also influence the coverage, our future work may include discussions of these design parameters as well.

\section{Acknowledgement}

This work was partly supported by the China Major State Basic Research Development Program (973 Prog-ram) No.2012CB316100(2), National Natural Science Foundation of China(NSFC) No.61171064, the China National Science and Technology Major Project No.2010ZX03003-003 and NSFC No. 61021001, where Ke xiong's work is  supported by NSFC No. 61201203.

\bibliographystyle{ieeetr}
\bibliography{basestation_analysis}

\end{document}